\documentclass[nohyper,pdftex]{JHEP3}
\pdfoutput=1
\usepackage{amsmath,amssymb}
\usepackage{graphicx}
\usepackage{cite}

\newcommand{\be}{\begin{equation}}
\newcommand{\ee}{\end{equation}}
\newcommand{\bea}{\begin{eqnarray}}
\newcommand{\eea}{\end{eqnarray}}
\newcommand{\eq}{\begin{equation}}
\newcommand{\eqx}{\end{equation}}
\newcommand{\eqn}{\begin{eqnarray}}
\newcommand{\bi}{\begin{itemize}}
\newcommand{\eqnx}{\end{eqnarray}}
\newcommand{\ei}{\end{itemize}}
\newcounter{hran}

\newcommand{\ba}{\begin{array}}
\newcommand{\ea}{\end{array}}
\newcommand{\balg}{\begin{align}}
\newcommand{\ealg}{\end{align}}

\newcommand{\lsim}
{\raise0.3ex\hbox{$\;<$\kern-0.75em\raise-1.1ex\hbox{$\sim\;$}}}
\newcommand{\gsim}
{\raise0.3ex\hbox{$\;>$\kern-0.75em\raise-1.1ex\hbox{$\sim\;$}}}

\title{A Minimal Inflation Scenario}
\author{Luis \'Alvarez-Gaum\'e$^{1}$,
C\'esar G\'omez$^{2}$, 
Raul Jimenez$^{3}$\\
$^{1}$ Theory Group, Physics Department, CERN, CH-1211, Geneva 23, Switzerland \\
$^{2}$ Physics Department and Instituto de Fisica Teorica UAM/CSIC, 28049 Cantoblanco, Madrid, Spain \\
$^{3}$ ICREA \& ICC, University of Barcelona (IEEC-UB), Marti i Franques 1, Barcelona 08028, Spain}

\abstract{We elaborate on a minimal inflation scenario based entirely on the general properties
of supersymmetry breaking in supergravity models. We identify the inflaton as the scalar component of the Goldstino superfield.  We write plausible candidates for the effective action describing this chiral superfield. In particular the theory depends (apart from parameters of $O(1)$) on a single free 
parameter: the scale of supersymmetry breaking.  This can be fixed 
using the amplitude of CMB cosmological perturbations and we 
therefore obtain the scale of supersymmetry breaking to be $10^{12-14}$ GeV. The model also incorporates explicit R-symmetry breaking in order to satisfy the slow roll conditions.  In our model the $\eta-$problem is solved without extra fine-tuning. We try to obtain as much information as possible in a model independent way using general symmetry properties of the theory's effective action, this leads to a new proposal on how to exit the inflationary phase and reheat the Universe.}

\begin{document} 
%%%%%%%%%%%%%%%%%%%%%%%%%%%%%%%%%%%%%%%%%%%%%%%%%%%%%%%%%%%%
\section{Introduction and summary of results}
%%%%%%%%%%%%%%%%%%%%%%%%%%%%%%%%%%%%%%%%%%%%%%%%%%%%%%%%
The inflationary paradigm provides a robust framework 
to explain the size, flatness, homogeneity of the universe and its 
perturbations \cite{guth:1981,mukhanov:1981,sato:1981,
albrecht.steinhardt:1982,guth/pi:1982,hawking:1982,linde:1982,starobinsky:1982,
bardeen/steinhardt/turner:1983}. It postulates that the universe went through a quasi de-Sitter 
phase in the past leading to an exponential expansion of space-time.  If this quasi de-Sitter 
phase is associated to a scalar field, then it can be shown, rather generally, that vacuum
fluctuations of the field are stretched by inflation to produce a nearly scale invariant power 
spectrum and with sufficient strength to produce the currently observed large scale structure in the 
universe \cite{mukhanov:1981,bardeen/steinhardt/turner:1983}. In spite of its successes, we still face
several questions to be answered in inflation, namely: what is the inflaton? how do we bring
the universe out of the inflationary phase? how can we stop inflation and transit to the 
decelerating/accelerating universe we live in today? what sets the energy scale of inflation? At a more fundamental level we have to deal with the problem of explaining the rich dynamics of the inflaton without too many free parameters and fine tuning.

In a previous letter \cite{AlvarezGaume:2010rt}, guided by the idea that the inflaton should 
be found naturally among the fields of any fundamental physics model, 
we have exploited some generic features of the supersymmetry 
(supergravity) breaking mechanism to design a model of inflation. 
In this model we identified the inflaton with
the order parameter of supersymmetry breaking and associated the supersymmetry breaking scale with that generating
cosmological perturbations. Under these conditions, and imposing explicit R-symmetry breaking, we showed 
that one can obtain enough number of e-foldings ($> 70$) to explain the observed universe.  Supersymmetry 
and inflation have a long history \cite{ellisetal, ovrutetal, rossetal, 
dvalietal, randalletal, lythriotto}, however in our approach we try to avoid making specific
and concrete models and try to see to what extent generic features of supersymmetry breaking
are enough to provide a good inflationary scenario.

Our main motivation to propose to identify the inflaton field with the order parameter of supersymmetry 
breaking is guided by the fact that, independently of the particular microscopic 
mechanism driving supersymmetry breaking (in what follows we will restrict ourselves to $F$-breaking ) we 
can, whenever we have violation of conformal invariance in the UV, define a superfield $X$ whose $
\theta$ component at large distances becomes the "Goldstino"
 (see \cite{volkovakulovrocek,seiberg2}). 
This superfield is the chiral superfield that appears in the divergence of the superfield of currents,
the Ferrara-Zumino (FZ) super-multiplet \cite{ferrarazumino} with universal properties at low energy
shared by large classes of models with supersymmetry breaking.  In the UV the scalar component $x$ of $X$ 
is well defined as a 
fundamental field while in the IR, once supersymmetry is spontaneously broken, this scalar field becomes the 
superpartner of the Goldstino i.e a two Goldstino state. The realization of $x$ as $GG$ can be 
implemented by imposing a non linear constraint in the IR for the $X$ field of the type 
$X^2=0$. In our previous approach to inflation we used one real component of the UV $x$ field as the
inflaton. We assumed the existence of a F-breaking effective superpotential for the $X$-superfield 
and we induced a potential for $x$ from gravitational corrections to the K\"ahler potential\footnote{Like most inflationary
theories containing supersymmetry, we present a simple model of multifield
inflation (sometimes called hybrid) \cite{lindehybrid}}.

In this paper we refine our model \cite{AlvarezGaume:2010rt} by exploring a family of K\"ahler
potentials depending on few parameters and which can lead to a reasonable cosmology.  In particular
we consider a simple class of models that apart from a few parameters of $O(1)$, depend on
the scale of supersymmetry breaking $f$.  The K\"ahler potential explicitly breaks the R-symmetry
as is needed in supergravity in order to have a slow-roll phase where the universe
inflates.  By fitting these models to cosmological data, we can read a supersymmetry breaking
scale of $O(10^{12-13})$ leading to a fairly heavy gravitino.  Given the scarcity of parameters in
the model presented, it is quite remarkable that many cosmological constraints can be satisfied
in such an economical manner.  The minimal choice we make has to also provide a graceful exit
from inflation without invoking a ``waterfall" field that will bring the theory out of exponential expansion.
In our case, the end of inflation is reached when the $X$-field begins to enter the nonlinear
phase ($X^2\,\sim\, 0$), the scalar component of the FZ chiral superfield is converted into
a pair of Goldstinos and the state of the universe can in principle be viewed as some weakly
interacting Fermi liquid.  The Fermi nature of the elementary components of the liquid creates
the necessary pressure to exit the inflationary period.  This is admittedly a far-off idea,
and we are currently exploring its microscopic properties in detail. We expect to report on
our results in a future publication \cite{landauandus}.  This detailed physical description
is not necessary in order to explore some of the phenomenological properties of our model.
This is what we will do in the rest of the paper.

Let us insist once more that the key feature of our philosophy is to obtain the basic properties
of the inflation scenario (inflation, graceful exit and reheating) out of a single superfield $X$
and the general properties of supersymmetry breaking.  

\section{Some motivation and the minimal model}

Our inspiration for the effective model we describe in this paper was motivated by some novel approaches
to inflation based on Higgs field and non-minimal supergravity couplings.
In reference \cite{ShaHiggs} (see also \cite{Barbon:2009ya}) an approach to inflation 
was suggested based on the standard
 model Higgs potential with the Higgs field coupled to gravity in a suitable Jordan frame. In this approach
the role of the inflaton is played by the Higgs field and the essential features of inflation appear as a
consequence of the chosen frame. The lagrangian for the Higgs
field in the Jordan frame is
\begin{equation}
L = \sqrt{-g_{J}} (R \Phi(h) - L(h;v))
\end{equation}
with $L(h;v)$ the standard Higgs lagrangian with $v$ the vev and $\Phi(h)$ a function of the Higgs 
field defining the frame. In Einstein frame the potential is given by
\begin{equation}
V_{E}(h) = \frac{V(h)}{\Phi(h)^2}
\end{equation}
with $V(h) = \lambda(h^2-v^2)^2$ the standard Higgs potential. For $\Phi(h) \sim (M_P^2+ \alpha h^2)$ we 
observe that the net effect of working in Jordan frame is to create an inflationary plateau for values of 
the Higgs field $h > \frac{1}{\alpha}$ in Planck units. In spite of its beauty and simplicity this approach suffers from several problems: fine tuning for $\alpha$, non unitary  higher order corrections etc. 
The study of ``Higgs inflation" was considered in supergravity with non-minimal
couplings in \cite{Einhorn:2009bh}.  The general approach using the full $N=1$ supergravity 
theory in the Jordan frame was worked out in  \cite{Ferrara:2010yw,Ferrara:2010in}.  See also in
this context \cite{Hertzberg:2010dc, Lee:2010hj, Kallosh:2010ug}.

We were inspired in the construction of our model by the Jordan frame formulation, and by some vague
similarities with no-scale supergravity theories \cite{noscalesugra}.  This is what led us to consider
together with the simplest form of a K\"ahler potential, a logarithmic correction.  This is suggested
by the form of the scalar potential in the Jordan frame (see \cite{Ferrara:2010yw} for details).
In the simplest case of $N=1$ supergravity coupled to a single chiral superfield $X$, the scalar
potential is determined completely by the K\"ahler potential $K(X,\bar X)$ and the superpotential 
$W(X)$ \cite{ferrarapotential}

\begin{equation}
V_{E} = e^{\frac{K}{M_P^2}}( -\frac{3}{M_P^2}W \bar W +G^{X,\bar X} D_XW \, D_{\bar X}\bar W),
\end{equation}
where the K\"ahler metric and the K\"ahler covariant derivatives are given by:
\begin{equation}
G_{X,\bar X}\,=\,\partial_X \bar \partial_X\, K(X,\bar X)\qquad
D\,W(X)\,=\,\partial_X\, W(X)\,+\,{1\over M_P^2}\,\partial_X\,K\, W(X)
\end{equation}
We now make an explicit choice for $K$ and $W$.

For us the inflaton superfield is the FZ-chiral superfield $X = z +\sqrt{2}\,\theta\psi + \theta^2 F$,
the order parameter of supersymmetry breaking. We will consider the simplest superpotential implementing
F-breaking of supersymmetry.  More elaborate superpotentials often reduce to this one once heavy fields
are integrated out.  In the following $M_P$ is the Planck mass.
\begin{equation}
W = f X + f_0\, M_P 
\end{equation}
with $f_0$ some constant to be fixed later by imposing the existence of a global minimum with 
vanishing cosmological constant and with $f$ the supersymmetry breaking scale $f= \mu_{susy}^2$.

As K\"ahler potential $K$ we consider
%\begin{equation}
% K= X \bar X + \frac{a}{M_P} X\,\bar X + \frac{b}{M_P^2}(X \bar X)^2
%+\frac{c}{M_P^2} (X^3 \bar X + c.c.)+\ldots\,-2\,M_P^2\,
%\log(1 + \frac{X +\bar X}{M_P})
%\end{equation}
\begin{equation}
K= X \bar X + \frac{a}{2 M_P} (X^2\,\bar X + c.c.) - \frac{b}{6 M_P^2}(X \bar X)^2
-\frac{c}{9 M_P^2} (X^3 \bar X + c.c.)+\ldots\,-2\,M_P^2\,
\log(1 + \frac{X +\bar X}{M_P})
\end{equation}
The coefficient $a$ will be determined by the condition that the minimum of the scalar potential
is at $z=0$.  With this choice, we can write explicitly (2.3) as a function
of $z,\bar z$ the scalar components of $X,\bar X$.  
After $f_0,a$ are determined, the condition
that the potential is a minimum at the origin imposes some constraints on the matrix of second
derivatives.  After some algebra we obtain the following results:
\begin{equation}
a=0\qquad f_0\,=\,-f\,M_P\qquad b+c-1 \geq 0 \qquad b-c-1\geq 0
\end{equation}
Using the real components of the $z$-field:
\begin{equation}
z\,=\,{1\over \sqrt{2}}\,(\alpha\,+\,i\,\beta),
\end{equation}
we can obtain their masses:
\begin{equation}
m^2_{\alpha}\,=\,\frac{f^2 (b+c-1)}{3 M^2},\qquad m^2_{\beta}\,=\,\frac{f^2 (b-c-1)}{3 M^2}.
\end{equation}
The potential is plotted in figure I for the limiting values $b=3$, $c=1$ .
It is easy to observe that the effect of the logarithm is to induce a waterfall at the end of inflation 
where reheating could take place.
Before entering to analyze the region where the slow roll conditions are achieved it is 
interesting to observe the form of the potential near the minimum $z=0$ in the limiting case
when $b\sim 1$ and $c\sim 0$.  In this case both scalar fields are nearly massless at the
origin, the potential is very flat there, and these fields could be used as a way of realising
a quintessence scenario.

Let us now compute the slow roll parameters $\epsilon$ and $\eta$. What we find, shown in the upper
panels of figure II,  is that slow roll inflation starts in a subplanckian region with 
fields of the order $0.1 M_P$ and ends for values of order $0.01 M_P$.

Pushing backwards from the end of inflation $70$ e-foldings we can compute the fluctuation spectrum 
in terms of the supersymmetry breaking scale $f$. This allows us to fix the value of $f$ in terms of 
the WMAP experimental data for the spectrum of fluctuations. This WMAP graph is presented 
in figure III. The value of $\mu_{susy}$ we get is $O(10^{13})$ GeV.  The dependence of $f$ on $b,c$ is rather
mild for $b,c$ values of $O(1)$.

Note that in the particular limiting case $b=1$,$\,c=0$, the K\"ahler potential is:
\begin{equation}
K= z\bar z + \frac{1}{M_P^2}(z \bar z)^2 -2M_P^2\,
\log(1 + \frac{z +\bar z}{M_P})
\end{equation}
and the R-symmetry breaking is all due to the logarithmic term.  This explicit breaking
is important to obtain the desired form of the potential and to satisfy the slow roll conditions.
If we ignore the polynomial terms in (2.10) and consider only the logarithmic part, then the
K\"ahler potential is invariant under imaginary shifts of $z$, this is analogous to what
happens in no-scale models and helps in generating the inflationary properties of our
potential. 

\begin{figure}
\begin{center}
\includegraphics[width=.45\columnwidth]{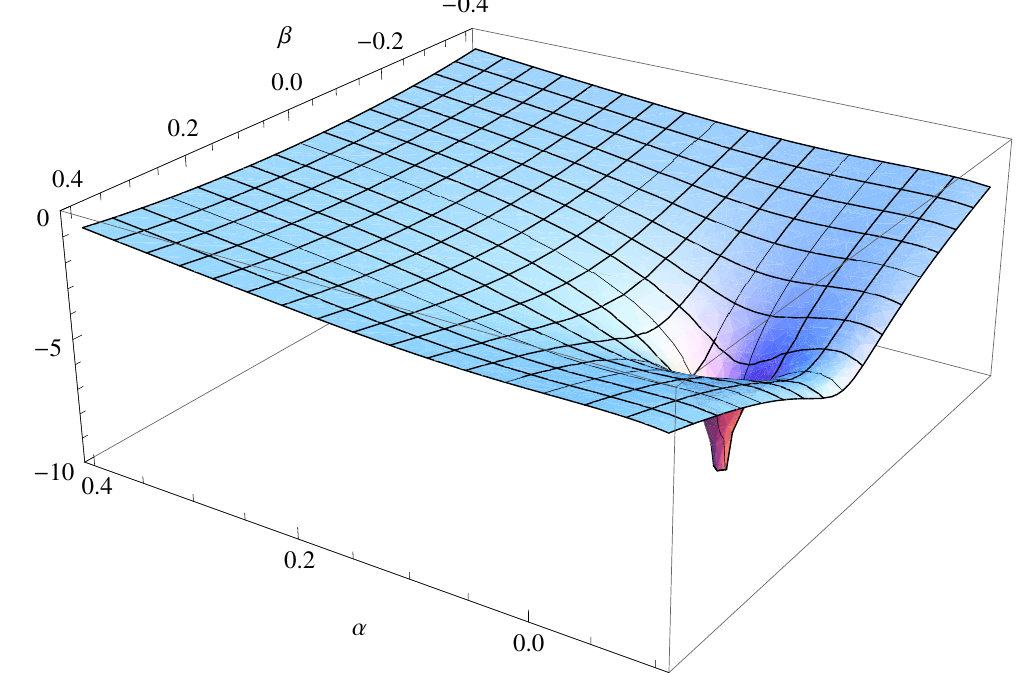}
\includegraphics[width=.45\columnwidth]{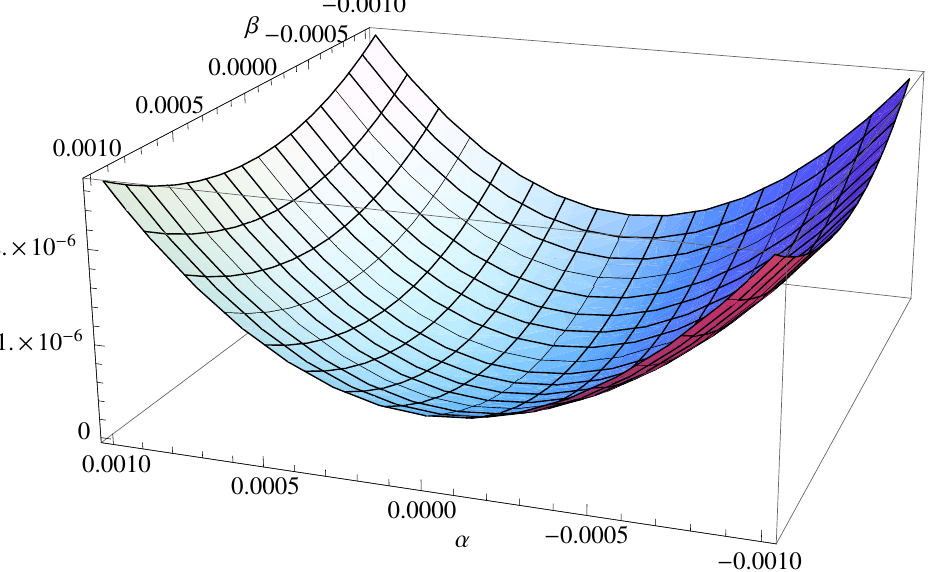}
\end{center}
\caption{The inflationary potential. Left panel: the shape of the potential, in logarithmic 
units of $M_{pl}^4$, for the two real fields $\alpha$ and $\beta$ and $b=3, c=1$.  Inflationary trayectories 
are obtained for all initial conditions of the field as it will always slow roll toward the 
steep throat located at $\alpha=\beta \sim 0$. Right panel: the minimum of the potential. 
This time the potential is shown in linear scale. Note that the potential is very flat and 
can therefore provide a natural candidate for quintessence.}
\label{fig:pot}
\end{figure}

\begin{figure}
\begin{center}
\includegraphics[width=.45\columnwidth]{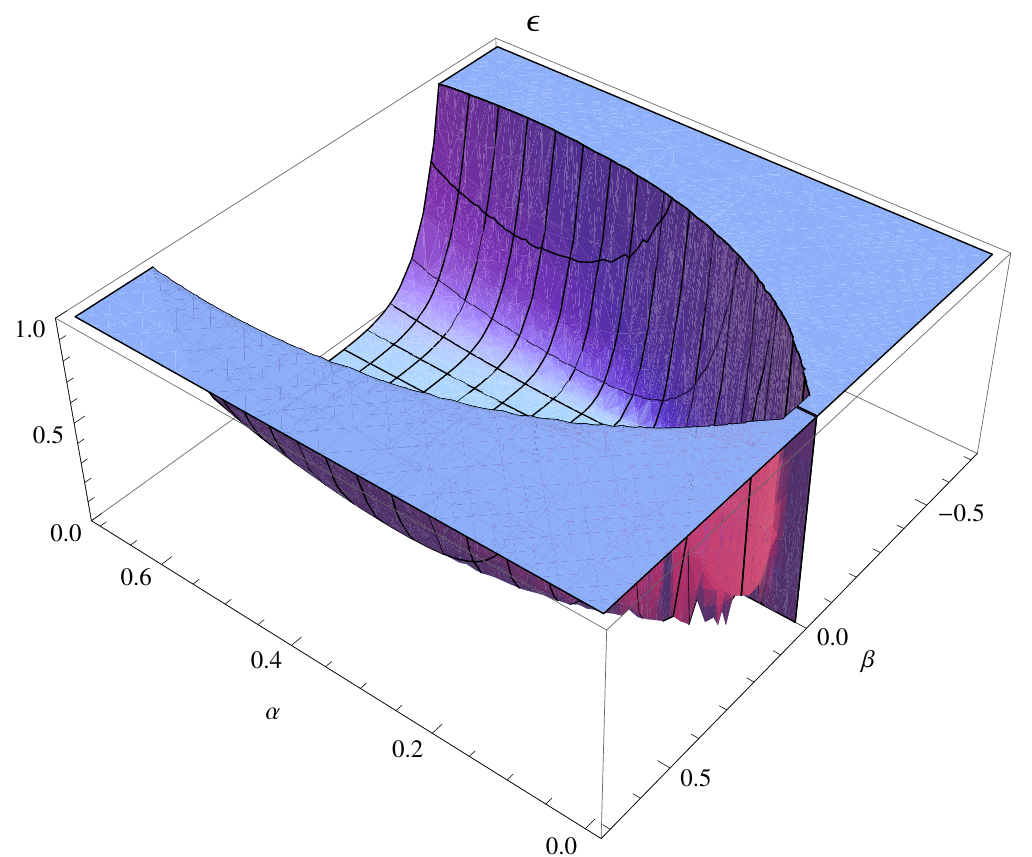}
\includegraphics[width=.45\columnwidth]{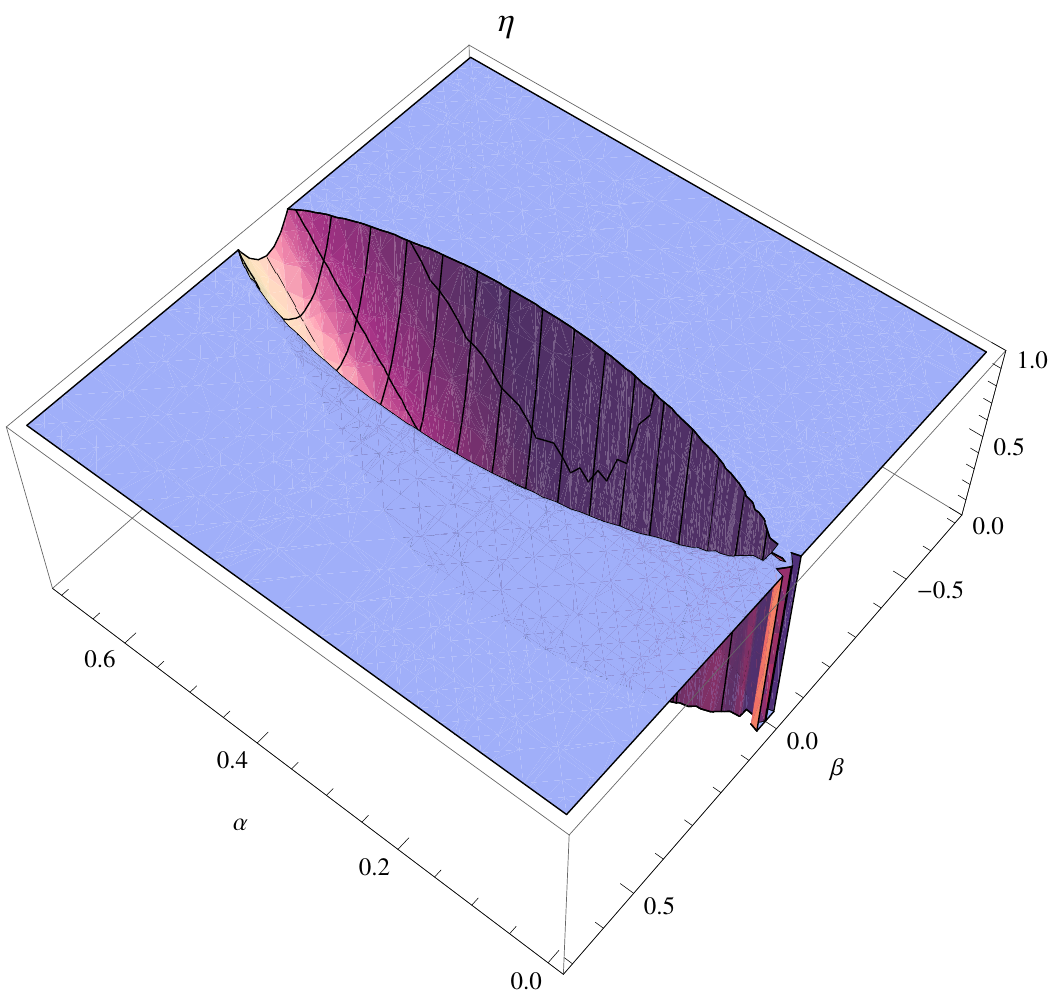}
\includegraphics[width=.45\columnwidth]{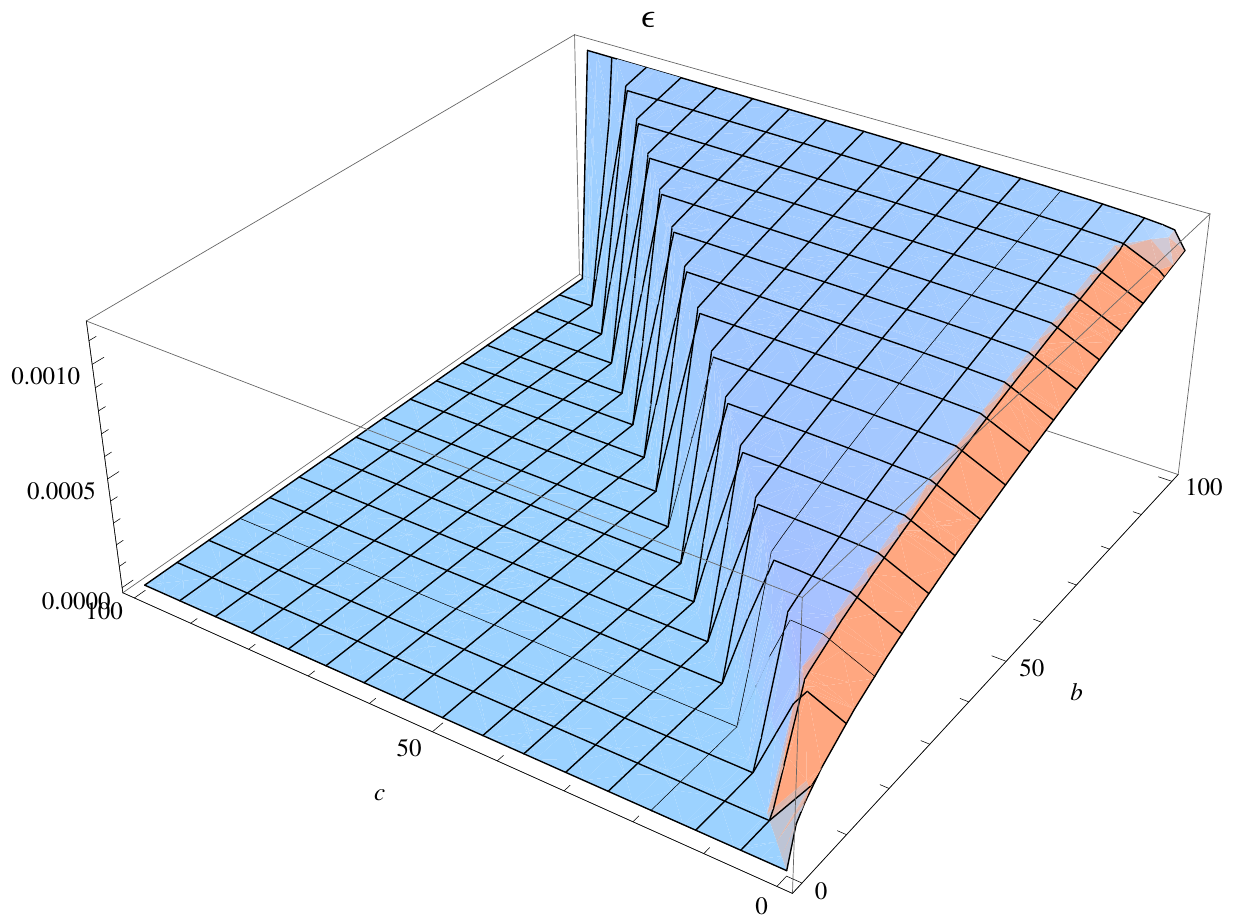}
\includegraphics[width=.45\columnwidth]{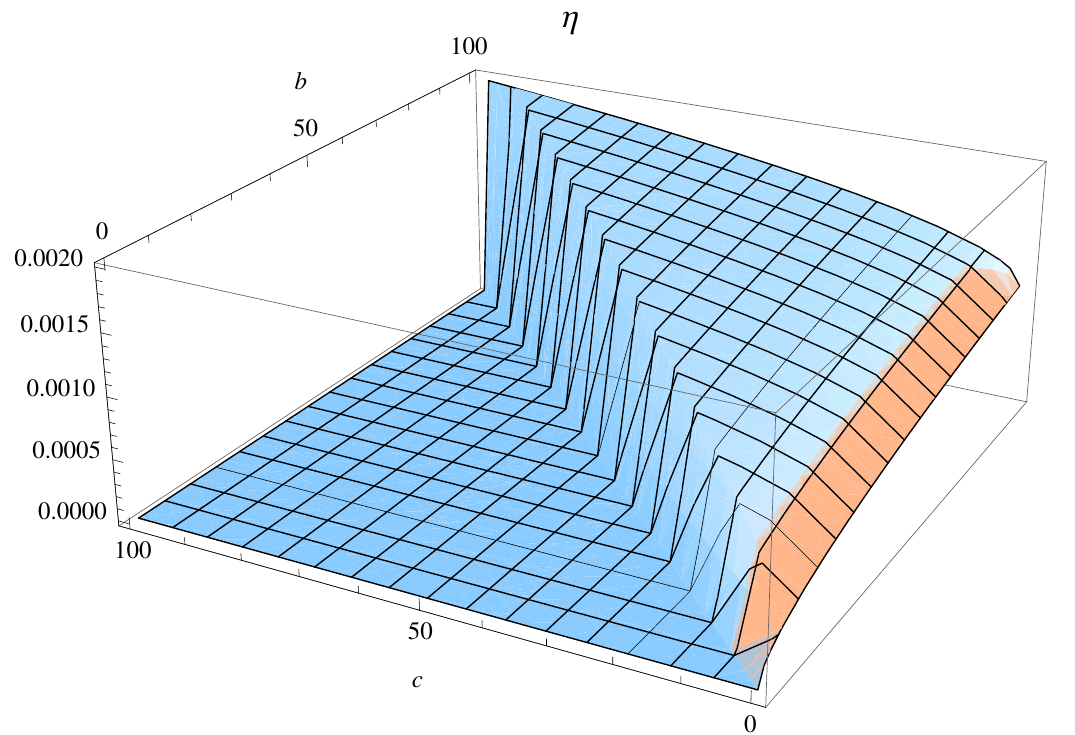}
\end{center}
\caption{In the two upper panels we plot the slow-roll parameters $\epsilon, \eta$ as a function of
 both fields, for the case $b=1, c=0$, to show that there are large regions where both $\epsilon$ and 
 $\eta$ are smaller than 1 and inflation can take place. Note also that inflation ends at 
 $\alpha, \beta \sim 0.01$ and that because of $\eta$ inflation is always sub-Planckian.  In the 
 two lower panels we plot $\epsilon$ and $\eta$ as functions of the two parameters 
 $b,c$ in our potential for some typical values of $\alpha,\beta$ to show that slow-roll conditions are natural in our model.}
\label{fig:slowroll}
\end{figure}

\begin{figure}
\begin{center}
\includegraphics[width=.5\columnwidth]{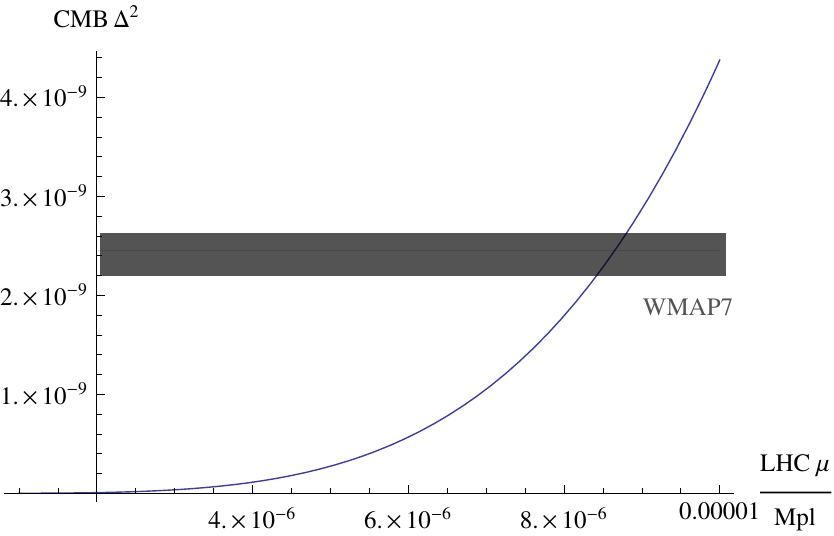}
\end{center}
\caption{Value of the supersymmetry breaking scale $\mu$ for which the amplitud of the scalar 
fluctuations, as measured in the CMB, are matched for our minimal model with $b=1, c=0$.}
\label{fig:susycmb}
\end{figure}

\section{Phenomenology of the model}

For the above model, leaving only as a free parameter 
the scale of supersymmetry breaking, we are now in a  
position to study its phenomenology. Recall that current observations of the cosmic microwave 
background (CMB) provide constraints on the amplitude of the fluctuations, the slope of the 
primordial power spectrum, upper limits to the amplitud of gravitational waves and the 
minimum number of required e-foldings. 

For simplicity we will focus in this section on particular values of $b, c$,  although 
any other allowed values of $b, c$ are equally easy to analyse and also provide acceptable 
fits to cosmological data. Actually it is easy to check that the value of the supersymmetry breaking 
scale determined by the experimental data on the spectrum of fluctuations is almost independent 
of the value of $b$ and $c$. Around the line $b=1+c$ one of the two scalar components of $z$ is 
very light giving rise to some potential form of quintessence. Generically the region of 
initial conditions leading to inflation reduces when we go deep inside the allowed 
region of values in the $b,c$ plane. We can determine the range of values for $f$ by 
matching the predicted level of the fluctuation 70 e-foldings before the slow roll 
parameters $\epsilon$, $\eta$ are $\sim 1$. This will provide a value for  $\alpha$, $\beta$ (recall that $z = \alpha + i \beta$) at which the level of fluctuations will have to match the ones measured by WMAP7 \cite{wmap7}.  As we show below this only happens for a small range of values of $f$.

First, we look at the overall shape of the inflaton potential. This is shown in Fig.~\ref{fig:pot} for the case $b = 3, c = 1$. 
The left panel shows the overal shape of the inflaton for ranges of the real fields $\alpha, \beta$ 
below $M_{pl}$. Note the flatness  of the potential. The deep throat 
at $\alpha=\beta \sim 0$ is an attractor and for any initial value of the field, it will 
end up the slow-roll toward the origin. This throat provides the graceful exit from 
inflation. The right panel, shows the minimum  of the potential, note that it is 
definite positive and naturally provides a very flat direction, which could be a 
good candidate for quintessence.      
 
Fig.~\ref{fig:slowroll} shows the value of the slow-roll parameters as a function of the 
fields $\alpha, \beta$. Note that inflation ends for $\alpha, \beta =   0.01$ and that 70 
e-foldings occur for $\alpha, \beta \sim 0.1$. At this point the value of the spectral slope 
of the primordial spectrum of fluctuations is in the range $n_s = 0.95 - 1.0$. The predicted 
value of the scalar-to-tensor ratio is $r  < 2 \times 10^{-3}$, thus it could be measurable by future 
experiments \cite{raulcmb}.

The constraint on the scale of supersymmetry breaking using the amplitude of fluctuations of the 
CMB is shown in Fig.~\ref{fig:susycmb}. 

In our scenario reheating will take place as a consequence of the existence of a waterfall that 
opens at the end of the inflationary period. The peculiar features of our choice of the inflaton 
field as the bosonic component of the supersymmetry breaking order parameter comes from the natural 
conversion of this field in the IR into pairs of Goldstinos $z \sim \frac{GG}{f}$ \cite{volkovakulovrocek, seiberg2}. 
This transmutation of the inflaton field at low energies into pairs of Goldstinos allows us to 
model the final state of the universe at the end of inflation as some sort of Fermi 
liquid with Fermi momentum
$p_{F} \sim \mu_{susy}$
in the simplest Fermi gas approximation. This Fermi gas gives us the first 
approximation to the positive pressure FRW phase after inflation.
The reheating temperature we get in this "barebone" approximation is
$T_{rh} \sim 10^9 - 10^{11} Gev$, and entropy production of the order of one.
More details on the graceful exit provided by the transmutation of the inflaton field into Goldstino
pairs will be given in a separate publication \cite{landauandus}.

Another interesting consequence of the Goldstino transmutation of the inflaton appears when we 
consider fluctuations at the end of the slow roll period of inflation. In this regime the 
spectrum of fluctuations for the composite $GG$ field goes as $k^3$ that is quite irrelevant 
in the IR \cite{Lyth}. Therefore, fluctuations produced during this phase, and any isocurvature mode, will be at very small scales and thus swamped by non-linearities caused by gravitational collapse in the decelerating FRW stage.

%
%%%%%%%%%%%%%%%%%%%%%%%%%%%%%%%%%%%%%%%%%%%%%%%%%%%%%%%%%%%%%%%%%%%%%%
%
\end{document}